\documentclass[aps,11pt]{revtex4}
\usepackage{amsmath, amsthm,times}
\usepackage{epsfig}
\usepackage{amssymb}
\usepackage{graphicx}
\usepackage{color}

\begin{document}

\sloppy

\title{Quantum Measurement Reliability versus Reversibility}

\author{S.\ J.\ van Enk and M.\ G.\ Raymer\footnote{Both authors contributed equally little to this work.}}

\address{
Department of Physics and Oregon Center for Optics\\ University of Oregon
\\
 Eugene, OR 97403}
\begin{abstract}
There is a constraining relation between the reliability of a quantum measurement and the extent to which the measurement process is, in principle, reversible. The greater the information that is gained, the less reversible the measurement dynamics become. To illustrate this relation, we develop a simple physical model for quantum measurement, as well as a hypothetical scheme by which the experimenters can determine the reliability and reversibility. We derive an ``uncertainty'' (constraining) relation between reliability and reversibility, which holds even when there is no interaction with any external environment other than the fundamental information recording device.  
\end{abstract}
\maketitle
\section{Introduction}
A tension between the ideas of complete human knowledge (information)
and the reversibility of physical dynamics played an important role in
the development of the concepts of probability and the predictability
(or lack thereof) of physical systems. To Pierre-Simon Laplace \cite{laplace}, in around 1800, the
world was fully deterministic, and ``Given an intelligence which could
comprehend all [...] for it, nothing would be uncertain, and the future,
as the past, would be present to its eyes.'' For Laplace, there were no
nature-imposed limits to gaining complete information or to solving
perfectly NewtonÕs equations to predict future events. Only our human
ignorance prevented us from doing so. ``Probability is relative, in
part to this ignorance, in part to our knowledge,'' he wrote.

Henri Poincar\'e \cite{poincare}, in the late 1800s,  argued instead that while there are no physical
limits to gaining information, there are physical limits to solving
NewtonÕs equations accurately to predict future events. He wrote that
``chance must be other than the name we give our ignorance [...] It may
happen that slight differences in the initial conditions produce very
great differences in the final phenomena; prediction becomes
impossible and we have the fortuitous phenomenon.''
Poincar\'e was the
first to get a glimpse of deterministic chaos in Newtonian physics---the
exponential sensitivity of outcomes to tiny changes of initial
conditions. He also stressed the significance of being able to reverse
(hypothetically) a systemÕs dynamics: ``The laws of nature bind the
antecedent [past] to the consequent [future] in such a way that the
antecedent is as well determined by the consequent as the consequent
by the antecedent.'' Given a tiny amount of ignorance about the system,
chaos would prevent one from perfectly reversing dynamics, even in principle.
Therefore, according to Poincar\'e, the absence of perfect reversibility is related to the need
to use probability theory to describe experimental outcomes.

Quantum theory changed all of that, of course, because now we know
there do exist physical limits to gaining information. The quantum
state of an object cannot be determined by any observations on that
single object \cite{DAriano,Alter}. Can we find a useful role for the idea
of dynamical reversibility in the context of quantum physics similar
to Poincar\'e's use in classical physics?
In the insightful 1965 paper ``Take a Photon'' \cite{frisch} by Otto Frisch,
a measurement on a single photon traversing an interferometer is discussed in the form of a comedic dialogue between various fictitious persona. One character  makes the statement ``Irreversibility is the very essence of information [...] To measure
 is to create information; and information is a state---in a machine
 or an organism---which extends from a certain time into the future.'' Before that, Brillouin in his book \cite{bril} from 1956  had stressed the fundamental relations between observations and irreversibility, titling one of the sections ``An Observation is an Irreversible Process.'' Brillouin in his turn cites von Neumann \cite{vN} as a source of many examples illustrating tradeoff relations between measurements and irreversibility in quantum mechanics. This reasoning goes back to
 Niels Bohr \cite{bohr} and his famous ``A phenomenon is not yet a phenomenon until
 it has been brought to a close by an irreversible act of
 amplification,'' as creatively  quoted by John
 Wheeler \cite{wheeler}. 
 
In much more recent times more precise quantitative bounds on information gain versus reversibility have been derived \cite{ban,maccone,buscemi}. Most studies use entropic measures of information and irreversibility, and go to great lengths to ensure that the most general type of quantum measurements is included in the description. Similarly, the closely related concept of information gain versus disturbance was studied quantitatively a decade ago \cite{fuchs}, again making sure bounds are valid for general classes of measurements, and many papers on the same subject have appeared since, see for example \cite{ozawa}.

Here our aim is more modest: we will not derive generally applicable bounds on information gain versus reversibility, but focus on one illustrative example that tries to be both realistic (at least as far as the measurement process is concerned) and simple.  The quantities describing reversibility and information we will use, neither of which are entropic quantities, arise naturally in a specific scenario in which our measurement model is used. Our model contains an adjustable parameter $ \eta $ that characterizes the strength and therefore the reliability of the measurement, with $\eta=0$ corresponding to no information gain, whereas $\eta=1$ corresponds to maximum reliability.
To
highlight the tradeoff we are interested in, our model contains no
decoherence other than that caused by the recording of the measurement
outcome itself. 
We thus find there exists a quantitative tradeoff between the reliability
of a measurement that is made and the ability to reverse fully the
measurement dynamics (in principle, if not in practice). 
\section{Measurement model}
We will describe here a measurement on a qubit, i.e., a quantum-mechanical two-state  system.
We denote a pure state of a qubit $q$ by
\begin{equation}\label{1}
|\psi\rangle_q=a|\downarrow\rangle_q+b |\uparrow\rangle_q,
\end{equation}
borrowing notation for the spin of spin-1/2 particles, where $a,b$ are complex coefficients satisfying $|a|^2+|b|^2=1$.
We wish to model a measurement on this qubit in the $|\uparrow\rangle$, $|\downarrow\rangle$ basis, also called ``spin up'' and ``spin down.''
\subsection{Unitary evolution}
We model the measurement as a two-step process, a unitary pre-measurement \cite{peres} and a final information-gathering step, which is not unitary \cite{vN}.  We split the pre-measurement stage into two steps: first we assume the qubit interacts with $N$  three-state systems (qutrits), where $N\gg 1$, according to the unitary transformation $\hat{U}$:
\begin{eqnarray}
|\downarrow\rangle_q\otimes |r\rangle_t^{\otimes N} &\stackrel{\hat{U}}{\mapsto}&
|\downarrow\rangle_q\otimes |d\rangle_t^{\otimes N}\nonumber\\
|\uparrow\rangle_q\otimes |r\rangle_t^{\otimes N} &\stackrel{\hat{U}}{\mapsto}&
|\uparrow\rangle_q\otimes |u\rangle_t^{\otimes N}.
\end{eqnarray}
The notation $|r\rangle_t^{\otimes N}$ means a product of $N$ kets of identical form, one for each qutrit. The state $|r\rangle_t$ is the initial ``ready" state, and $|d\rangle_t$ and $|u\rangle_t$ are states which become associated with the qubit states ``spin down'' and ``spin up'' by $\hat{U}$.  This step can be viewed as an amplification process. One can think of an avalanche photo detector or a bubble chamber particle detector, where one photon or one particle can create a state change in many other particles: one can think of each qutrit as being in a metastable ``ready'' state \cite{vK}, $|r\rangle_t$, with two possible paths to decay to a lower-lying stable state, either $|u\rangle_t$ or $|d\rangle_t$, triggered by the qubit. (A simple classical analogy would be a domino standing on end, which can be knocked over in the forward or backward direction by a small kick.)

In the next step the signal is converted to a macroscopic signal: the qutrits are coupled to one degree of freedom of a macroscopic system (the ``counter"). The  counter is modeled as a harmonic oscillator (a pendulum, for instance) with resonance frequency $\omega$. We assume it starts off in the ground state $|0\rangle_c$ , that is, the minimum uncertainty wavepacket (coherent state) with amplitude 0.  
The  unitary transformation $\hat{U}'$ acting on  each qutrit and the harmonic oscillator is assumed to be
\begin{eqnarray}
|u\rangle_t\otimes |\alpha\rangle_c&\stackrel{\hat{U}'}{\mapsto}&
|u\rangle_t \otimes \hat{D}(\epsilon)|\alpha\rangle_c\nonumber\\
|d\rangle_t\otimes |\alpha\rangle_c&\stackrel{\hat{U}'}{\mapsto}&
|d\rangle_t\otimes \hat{D}(-\epsilon)|\alpha\rangle_c\nonumber\\
|r\rangle_t\otimes |\alpha\rangle_c&\stackrel{\hat{U}'}{\mapsto}&
|r\rangle_t \otimes |\alpha\rangle_c,
\end{eqnarray}
where $|\alpha\rangle_c$ denotes a minimum uncertainty wavepacket (coherent state) of the counter with amplitude $\alpha$,  and the displacement operators $\hat{D}(\pm\epsilon)$ act on the wavepacket as
\begin{equation}
\hat{D}(\pm \epsilon)|\alpha\rangle_c=|\alpha\pm \epsilon\rangle_c.
\end{equation}
The $N$ qutrits together thus interact with the counter oscillator through the unitary transformation $\hat{U}''\equiv\hat{U}'^{\otimes N}$:
\begin{eqnarray}
|u\rangle_t^{\otimes N} |0\rangle_c&\stackrel{\hat{U}''}{\mapsto}&
|u\rangle_t^{\otimes N} |N\epsilon\rangle_c\nonumber\\
|d\rangle_t^{\otimes N} |0\rangle_c&\stackrel{\hat{U}''}{\mapsto}&
|d\rangle_t^{\otimes N} |-N\epsilon\rangle_c.
\end{eqnarray}
The displacement operator simply translates the oscillator in position by an amount proportional to $\alpha(0)$, after which the oscillator will, indeed, oscillate at frequency $\omega$: the amplitude as a function of time will be $\alpha(t)=\alpha(0)\exp(-i\omega t)$, where $\alpha(0)=\pm N\epsilon$ in our case.
 
Summarizing, starting with a qubit in the pure state (\ref{1}), after the pre-measurement the final state  of the qubit + detector atoms + counter would be
\begin{equation}
|\Psi\rangle_{q,t,c}=a|\downarrow\rangle_q|d\rangle_t^{\otimes N}|-N\epsilon\rangle_c+b |\uparrow\rangle_q|u\rangle_t^{\otimes N}|+N\epsilon\rangle_c.
\end{equation}

\begin{figure}
\includegraphics[width=4.5in]{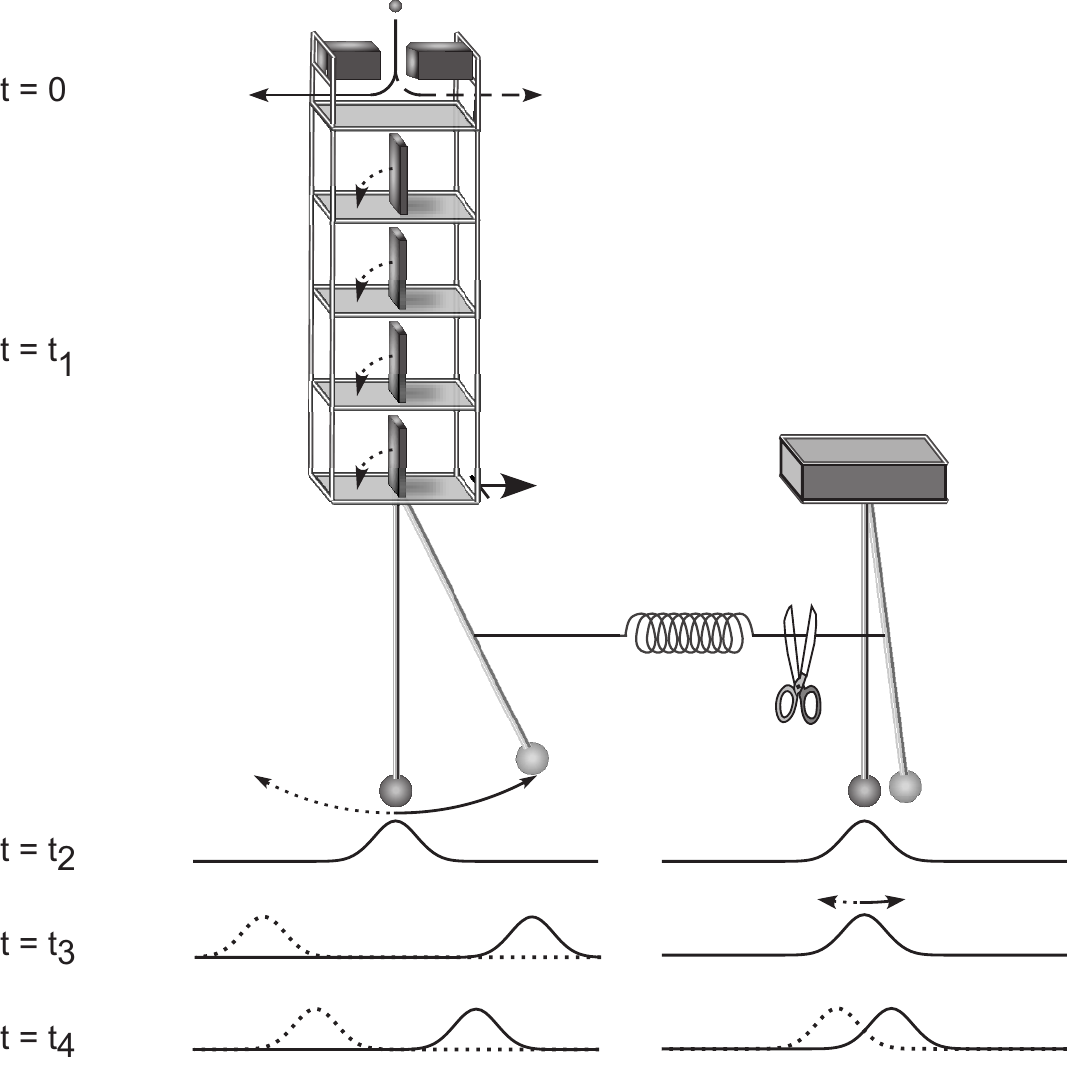}
  \caption{Gedanken experiment to illustrate the measurement model. At time $t=0$ the qubit, modeled as a particle with two spin states, drops through an inhomogeneous magnetic field, like in a Stern-Gerlach apparatus, upon which it gets deflected either rightward or leftward depending on its spin state. The magnets, being mounted on a movable carriage (with no frictional forces impeding it) recoils left or right by the principle of momentum conservation. The slight recoiling of the carriage is sufficient to cause all the dominos balanced on their ends to topple either leftward or rightward, all going the same direction at time $t=t_1$. The large amount of energy previously stored in the metastable dominos causes, at $t=t_2$, the carriage to recoil even more vigorously, which sets in motion a pendulum (the counter), which is attached to the bottom of the carriage. 
  We assume that the energy of the falling dominos is {\em not} dissipated away, in order to allow the possibility of complete reversibility in principle.
  In the measurement stage, a second pendulum (the ``probe'') is weakly coupled to the first pendulum at $t=t_3$, causing the probe to oscillate, but with a much smaller amplitude than the counter pendulum. At time $t=t_4$ the counter and probe are correlated, both having a positive (or negative) amplitude. }\label{domino}
 \end{figure}

These pre-measurement steps are illustrated in gedanken-fashion (not meant to be a physically realistic model) in Figure~\ref{domino}. 
\subsection{Information gathering}
The final step of the measurement consists of a {\em partial} or imperfect measurement on the counter state. This is modeled by ``splitting off'' a fraction $\eta$ of the coherent state of the counter and performing a projective measurement on that fraction. To do this, we introduce a second macroscopic oscillator, called the ``probe,'' also starting off in the ground state. Let $ \eta $ be the ``strength'' of the measurement, which obeys $0\leq \eta\leq 1$. 
At an appropriate time, such that $\exp(-i\omega t)=1$ and hence $\alpha(t)=\alpha(0)$,
 we apply the following transformation:
\begin{equation}
|\alpha(0)\rangle_c|0\rangle_p\mapsto|\sqrt{1-\eta}\alpha(0)\rangle_c|\sqrt{\eta}\alpha(0)\rangle_p,
\end{equation}
where, in our case, $\alpha(0)=\pm N\epsilon$. We can picture this transformation  as follows: the ``counter'' oscillator kicks the ``probe'' oscillator and transfers a fraction $\eta$ of its total energy, after which both pendulums oscillate, the counter now having smaller amplitude than initially. By this mechanism, the probe ends up in a linear combination of the state $|\sqrt{\eta} N\epsilon\rangle_p$ and the state $|-\sqrt{\eta} N\epsilon\rangle_p$, and is entangled with the qubit+qutrit+counter system. 

Then we assume we perform the {\em optimum} measurement on the probe to distinguish the two states $|\pm\sqrt{\eta} N\epsilon\rangle_p$. These two states are not orthogonal and so cannot be distinguished with certainty.
 \begin{figure}
\includegraphics[width=3.0in]{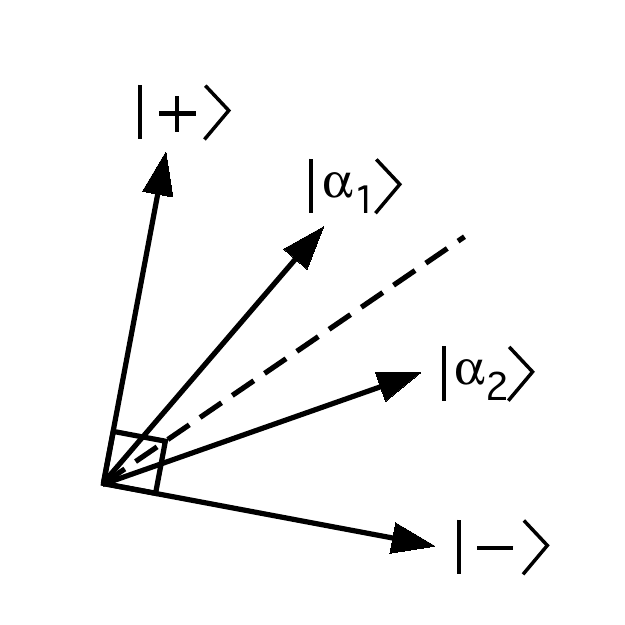}
  \caption{Illustration of the optimum measurement to distinguish two nonorthogonal (coherent) states $|\alpha_1\rangle$ and $|\alpha_2\rangle$, which are depicted here in Hilbert space. One projects onto two orthogonal states $|\pm\rangle$ that straddle symmetrically the states to be distinguished.}\label{Dol}
 \end{figure}
The optimum measurement, in the sense of maximizing the probability of a correct measurement outcome \cite{helstrom}, is illustrated in Fig.~\ref{Dol}, and projects onto the following two orthogonal states
\begin{equation}
|\pm\rangle_p=\gamma|\pm \sqrt{\eta} N\epsilon\rangle_p-\beta
|\mp \sqrt{\eta} N\epsilon\rangle_p,
\end{equation}
with
\begin{eqnarray}
\gamma=\frac{\sqrt{1+c(\eta)}+\sqrt{1-c(\eta)}}{2\sqrt{1-c(\eta)^2}},\nonumber\\
\beta=\frac{\sqrt{1+c(\eta)}-\sqrt{1-c(\eta)}}{2\sqrt{1-c(\eta)^2}},
\end{eqnarray}
where $c(\eta)$ is the overlap between the two states to be distinguished. We write $c(\eta)$ explicitly as a function of $\eta$ to emphasize that it is the principal adjustable parameter:
\begin{equation}
c(\eta)=\langle +\sqrt{\eta}N\epsilon|-\sqrt{\eta}N\epsilon\rangle=\exp(-2\eta N^2\epsilon^2).
\end{equation}
It is convenient to define the constant 
\begin{equation}
c_{0}=\langle +N\epsilon|-N\epsilon\rangle=\exp(-2 N^2\epsilon^2),
\end{equation}
which is the minimum value that $c(\eta)$ could assume for fixed values of $N$ and $\epsilon$, namely when $\eta=1$.
In terms of $c_0$ we have
\begin{equation}
c(\eta)=(c_0)^\eta.
\end{equation}
That is, the relevant parameter describing the measurement process  depends exponentially on the measurement strength $\eta$.

The probability of error, corresponding to projecting onto the ``wrong'' outcome $|\pm\rangle_p$ when the input state was $|\mp \sqrt{\eta}N\epsilon\rangle_p$, is
\begin{equation}\label{Perror}
P_{{\rm error}}=|\langle \mp|\pm \sqrt{\eta}N\epsilon\rangle|^2=\frac{1-\sqrt{1-c(\eta)^2}}{2}.
\end{equation}
In terms of the error probability and the coefficients $a$ and $b$ of our initial state, the
outcome ``$+$'' occurs with probability $P_+=(1-P_{{\rm error}})|b|^2+P_{{\rm error}}|a|^2$, and the outcome ``$-$'' occurs with probability $P_-=(1-P_{{\rm error}})|a|^2+P_{{\rm error}}|b|^2$.
In the limit of $\eta\rightarrow 1$ and $N\epsilon\rightarrow\infty$, we have $c(\eta)\rightarrow 0$, and
$|\pm\rangle_p\rightarrow |\pm \sqrt{\eta} N\epsilon\rangle_p$ and $P_{{\rm error}}\rightarrow 0$, thus describing an ideal von Neumann measurement. In the opposite limit $\eta\rightarrow 0$, we have $P_{{\rm error}}=1/2$, corresponding to no information gain at all (purely guessing a measurement outcome will yield an error probability of 50\% as well).

The final (unnormalized) state of qubit, the qutrits, and the counter, after measuring the probe is either
\begin{equation}\label{ab1}
a\sqrt{P_{{\rm error}}}|\downarrow\rangle_q
|d\rangle_t^{\otimes N}|-\sqrt{1-\eta}N\epsilon\rangle
+b\sqrt{1-P_{{\rm error}}}|\uparrow\rangle_q
|u\rangle_t^{\otimes N}|+\sqrt{1-\eta}N\epsilon\rangle,
\end{equation}
or
\begin{equation}\label{ab2}
a\sqrt{1-P_{{\rm error}}}|\downarrow\rangle_q
|d\rangle_t^{\otimes N}|-\sqrt{1-\eta}N\epsilon\rangle
+b\sqrt{P_{{\rm error}}}|\uparrow\rangle_q
|u\rangle_t^{\otimes N}|+\sqrt{1-\eta}N\epsilon\rangle,
\end{equation}
depending on the measurement outcome.

Note that we are using the standard quantum measurement hypothesis to describe the measurement of the probe. That is, we are not attempting to ``solve'' the ``quantum measurement problem,'' which is the question whether one should search for a physical mechanism by which quantum amplitudes are converted into experimental actualities \cite{legg}.

The above optimal measurement on the probe can in fact be performed when considering coherent states of light beams: the optimum measurement was first conceived by Dolinar \cite{dolinar} and later implemented in Ref.~\cite{jm}.

\subsection{Reversing evolution}
A key element of our treatment is to consider to what extent the measurement dynamics are reversible, and to ascertain what constraints are placed on this reversibility by virtue of leaving a permanent trace (information) in the probe. Therefore let us consider an idealized optimal reversal scenario, which serves to make our conclusions as clear and as simple as possible.

After the measurement on the probe, one could in principle perform a unitary operation \footnote{The unitary operator that enacts the unitary transformation (\ref{U3}) is explicitly given by  \[
 |\uparrow\rangle_q\langle \uparrow|\otimes |r\rangle_t\langle u|\otimes \hat{D}(-\sqrt{1-\eta}\epsilon)+ |\downarrow\rangle\langle \downarrow|\otimes |r\rangle\langle d|\otimes \hat{D}(+\sqrt{1-\eta}\epsilon)
 \]
 } (which depends on the value of $\eta$, but not on the measurement outcome) on the combined system of counter, all qutrits, and the qubit (but not the probe):
 \begin{eqnarray}\label{U3}
|\uparrow\rangle_q|u\rangle_t^{\otimes N}|+\sqrt{1-\eta}N\epsilon\rangle_c&\mapsto&
|\uparrow\rangle_q|r\rangle_t^{\otimes N}|0\rangle_c\nonumber\\
|\downarrow\rangle_q|d\rangle_t^{\otimes N}|-\sqrt{1-\eta}N\epsilon\rangle_c&\mapsto&
|\downarrow\rangle_q|r\rangle_t^{\otimes N}|0\rangle_c.
\end{eqnarray}
The qutrits and the counter are thus reset to their initial states, and the qubit $q$ thus completely disentangles from the qutrits and the counter.  This step is certainly far from realistic for photomultipliers, bubble chambers, and Geiger counters, but as a matter of principle the laws of Nature do allow this reversal. In a recent experiment \cite{undo} a partial reversal of a ``weak'' quantum measurement on a superconducting qubit was demonstrated.

In our gedanken experiment in Figure~\ref{domino}, one can imagine first cutting the coupling spring between probe and counter, then exerting a force on the counter pendulum just perfectly so that it comes to rest and all the dominos are flipped back into their upright positions. This will work perfectly only if the qubit is sent back into the magnetic field at the right moment so that it absorbs the momentum necessary to bring the carriage to rest while deflecting the qubit back upward.

Because the probe is now in a definite observed state, either $|+\rangle_p$ or $|-\rangle_p$, it also is not entangled with the qubit. In fact, the projection of the probe onto the states $|\pm\rangle_p$ 
implies a collapse of the qubit $q$ into two possible states as well \cite{collapse}:
for the result ``$+$'' the state of the qubit ({\em after} the reversal operation (\ref{U3}) and inserting the proper normalization factor) follows from Eq.~(\ref{ab1}),
\begin{equation}\label{plus}
|\psi_+\rangle_q=\frac{a\sqrt{P_{{\rm error}}}|\downarrow\rangle_q+b\sqrt{1-P_{{\rm error}}}|\uparrow\rangle_q}{\sqrt{P_+}},
\end{equation}
and for the result ``$-$'' the qubit is collapsed into the state
\begin{equation}\label{min}
|\psi_-\rangle_q=\frac{a\sqrt{1-P_{{\rm error}}}|\downarrow\rangle_q+b\sqrt{P_{{\rm error}}}|\uparrow\rangle_q}{\sqrt{P_-}}.
\end{equation}
In the case that $P_{{\rm error}}=1/2$, these two states are in fact the same, and equal to the initial qubit state: indeed, no measurement has been performed in this case, since $\eta=0$, and no collapse has taken place either after the reversal (\ref{U3}). In the other extreme limit, $P_{{\rm error}}=0$, the collapse is complete, and the qubit ends up either in the state $|\downarrow\rangle_q$ or
in the state $|\uparrow\rangle_q$, with probabilities $|a|^2$ and $|b|^2$, respectively, as it should according to standard textbook Quantum Mechanics.
\section{Reliability vs reversibility: a scenario}
In order to give an operational definition of reversibility and reliability, we consider a specific scenario.
We consider Alice and Bob: Alice prepares a qubit and hands it Bob. Bob will perform the measurement discussed in the preceding Section on that qubit, with the value of the measurement strength  $\eta$ chosen by Bob. After the measurement, Bob tries to reverse the evolution of the qubit and detector system as completely as he can [by using the $\eta$-dependent unitary operation (\ref{U3})], tells Alice the result of his measurement, and returns the qubit to Alice. She will then perform one of two measurements. Bob will not know in advance which of the two measurements Alice will perform. Either she will check whether Bob's measurement result produced reliable information for him, or she will check whether his measurement changed the quantum state that she had prepared. This procedure may be repeated multiple times, although this is not necessary.

In order to maintain a description using only pure states, we specify that in each run of the experiment, Alice prepares {\em two} qubits $A$ and $B$ in a maximally entangled state (one of the Bell states) of the form 
\begin{equation}
|\Psi\rangle_{A,B}=\frac{1}{\sqrt{2}}\left[|\uparrow\rangle_A|\uparrow\rangle_B+|\downarrow\rangle_A|\downarrow\rangle_B\right]\equiv|\Psi^+\rangle_{A,B}.
\end{equation}
Bob knows Alice prepares this state. Alice hands Bob the qubit $B$, but holds qubit $A$ in her possession, such that Bob has no access to it.

Bob has a device that implements both the measurement and the reverse evolution described in the preceding Section. The device has a knob and a button: the knob to set the value of $\eta$, and the button to reverse the evolution as completely as possible. The device does not attempt to apply a reversal operation to the probe (or to Bob's brain, which stores information about the probe), since Bob does not allow this
\footnote{Of course, assuming that Bob and his device form a closed system, an outside observer (Wigner, or his friend Szilard, presumably) may well decide to reverse completely the evolution of Bob's probe and brain, too, at least in principle. That observer, though, will not gain any information. We describe our scenario from the point of view of Bob and Alice only.}. The device sends qubit $B$ back to Alice, and Bob tells Alice of the result of his measurement, which is either ``spin up,'' or ``spin down,'' depending on which measurement outcome he obtained. That is, he assumes there is a correlation between ``spin up (down)'' and the probe state ``$+ (-)$.''

Upon receiving qubit $B$ from Bob,  Alice  now performs one of two measurements, which she chooses at random with 50\% probability each:
\begin{enumerate}
\item  She may measure qubit $A$ in the spin-up, spin-down basis,  performing a perfectly reliable ideal von Neumann measurement. This allows her to determine what spin value Bob should have measured ideally, because the initial entangled state  ensures that her spin has the same value as Bob's if they {\em both} perform {\em ideal} measurements in the same spin-up, spin-down basis.
\item She may perform an ideal standard (von Neumann) measurement on the two qubits $A$ and $B$ together, projecting onto the four Bell states:
\begin{eqnarray}
|\Psi^{\pm}\rangle_{A,B}&=&\frac{1}{\sqrt{2}}\left[|\uparrow\rangle_A|\uparrow\rangle_B\pm|\downarrow\rangle_A|\downarrow\rangle_B\right]\nonumber\\
|\Phi^{\pm}\rangle_{A,B}&=&\frac{1}{\sqrt{2}}\left[|\uparrow\rangle_A|\downarrow\rangle_B\pm|\downarrow\rangle_A|\uparrow\rangle_B\right]
\end{eqnarray}
In this case, she counts the result $|\Psi^+\rangle$ as a ``yes,'' and the remaining three outcomes as a ``no.''
\end{enumerate}
In case 1, she can compare her result with Bob's measurement outcome. The probability that her result differs from that of Bob's, that is that Bob's measurement is wrong, is simply given by $P_{{\rm error}}$ given in Eq.~(\ref{Perror}). 

In case 2, the probability of obtaining the ``yes'' outcome, indicating that the joint state of qubits $A$ and $B$ is the same as it was originally, is given by the ``fidelity'' $F$, which is defined as the overlap between the final and initial states of the two qubits together, averaged over Bob's two possible measurement outcomes. After Bob has obtained the ``$+$'' outcome, the joint state is
\begin{equation}
|\phi_+\rangle_{AB}=\sqrt{P_{{\rm error}}}|\downarrow\rangle_A|\downarrow\rangle_B+\sqrt{1-P_{{\rm error}}}||\uparrow\rangle_A\uparrow\rangle_B,
\end{equation}
which is similar to Eq. (\ref{plus}), whereas after the ``$-$'' outcome the state would be
\begin{equation}
|\phi_-\rangle_{AB}=\sqrt{1-P_{{\rm error}}}|\downarrow\rangle_A|\downarrow\rangle_B+\sqrt{P_{{\rm error}}}||\uparrow\rangle_A\uparrow\rangle_B.
\end{equation}
The overlaps are in fact the same in the two cases, and these define the fidelity,
\begin{equation}
F=|\langle \Psi^+|\phi_+\rangle|^2=|\langle \Psi^+|\phi_-\rangle|^2=\frac{1}{2}+\sqrt{P_{{\rm error}}(1-P_{{\rm error}})}.
\end{equation}
Rewriting the right-hand side as a function of $c(\eta)$ yields
  \begin{equation}\label{F}
 F=\frac{1+c(\eta)}{2}.
 \end{equation}
The fidelity is a measure of Bob's ability to reverse the qubit-plus-detector evolution, and is governed by the device setting $\eta$ that Bob chose. 

\begin{figure}
\includegraphics[width=4.5in]{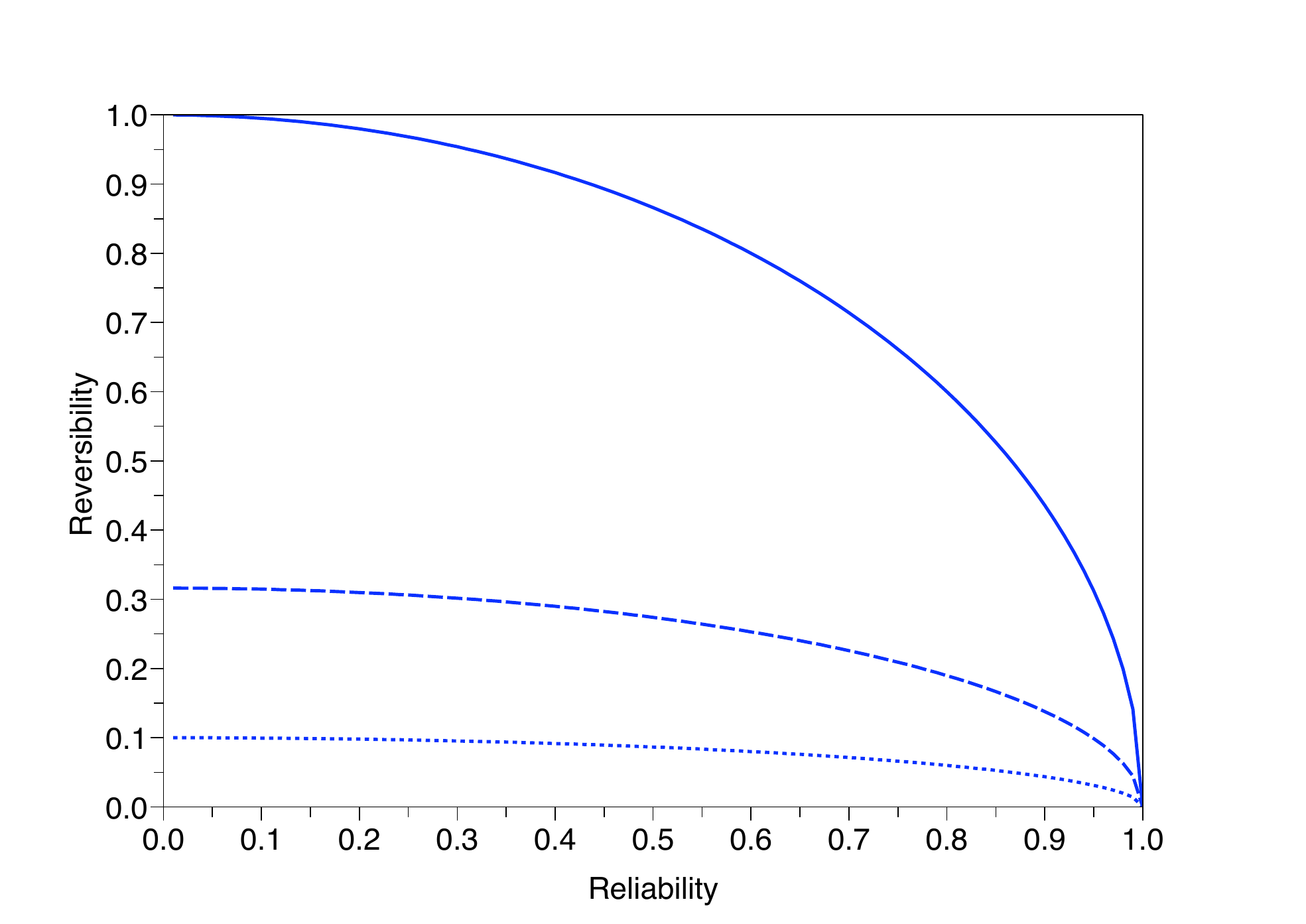}
  \caption{Degree of reversibility  $D_{{\rm rev}}$ versus degree of reliability
  $D_{{\rm rel}}$ for three values of the parameter $K: K=1,10,100$ (which characterizes the effect of other observers, see Eqs.~(\ref{K})--(\ref{K2}) for the solid,  dashed and dotted curves, respectively.}\label{rr}
 \end{figure}

From the two results (\ref{Perror}) and (\ref{F}) we obtain the tradeoff relation
 \begin{equation}\label{tradeoff}
 (2F-1)^2+(1-2P_{{\rm error}})^2=1.
 \end{equation}
 That is, Bob can increase his ability to reverse the qubit evolution by increasing the fidelity $F$,  but only at the cost of increasing the error probability of his measurement, and {\em vice versa}. This is our main result. 
The relation (\ref{tradeoff}) encourages us to define the following two quantities: a degree of reliability,
 \begin{equation}
D_{{\rm rel}}=1-2P_{{\rm error}}=\sqrt{1-c(\eta)^2}
\end{equation}
and a degree of reversibility:
 \begin{equation}
D_{{\rm rev}}=2F-1=c(\eta).
\end{equation}  
By virtue of equation (\ref{tradeoff}), these two quantities can alternatively be written in terms of an angle $\theta$, such that
\begin{eqnarray}
D_{{\rm rel}}&=&\sin\theta,\nonumber\\
D_{{\rm rev}}&=&\cos\theta.
\end{eqnarray}
In Fig.~\ref{rr} we plot as the solid curve $D_{{\rm rev}}$ vs $D_{{\rm rel}}$ (for a single observer, Bob).
 
\begin{figure}
\includegraphics[width=4.5in]{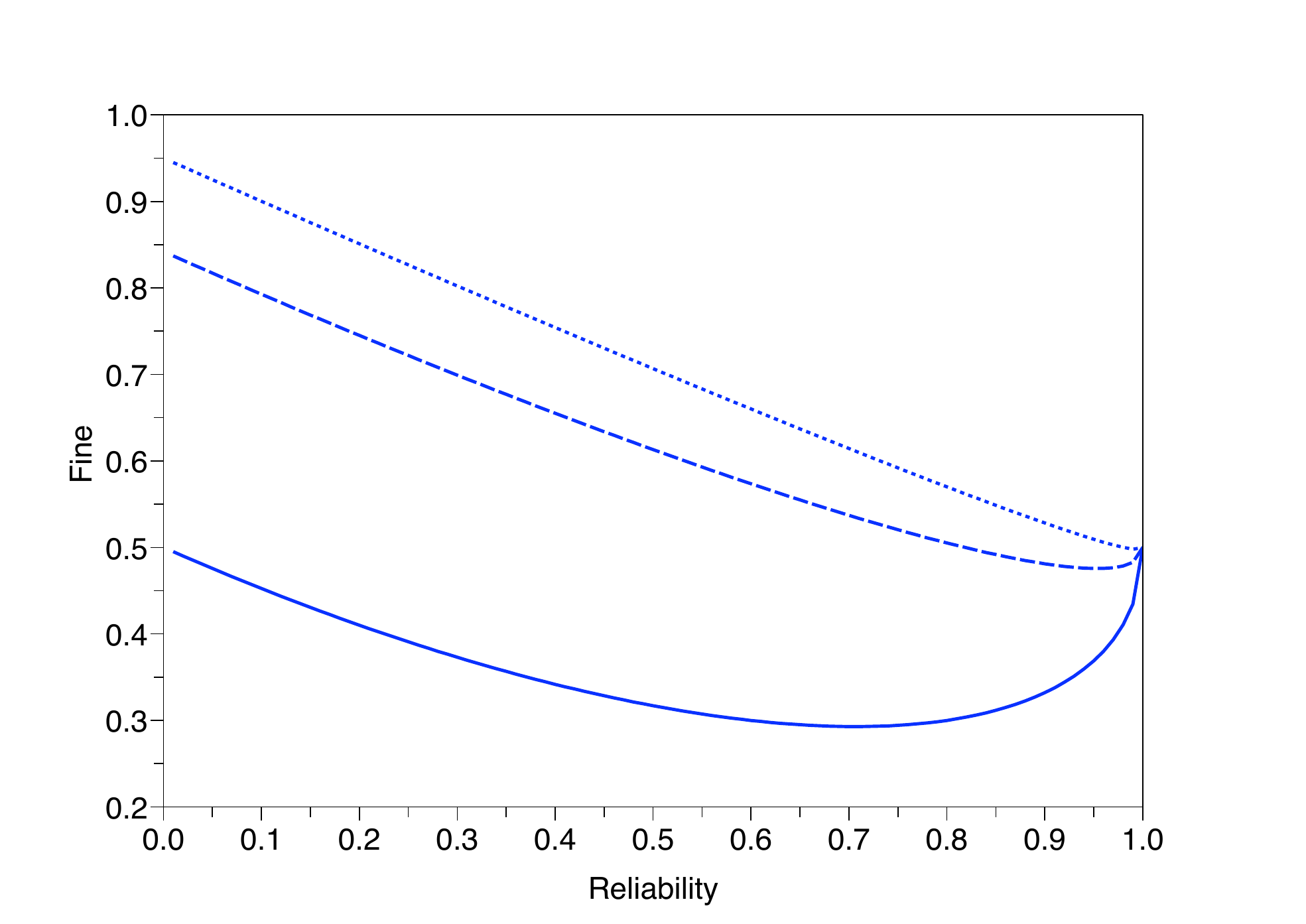}
  \caption{Bob's cost function to be minimized: the fine $\langle f\rangle$ as defined in Eq.~(\ref{f}), versus the degree of reliability
  $D_{{\rm rel}}$,   for the same three values of the parameter $K: K=1,10,100$ as in the previous figure.}\label{Fine}
 \end{figure}

Depending on which quantity as a function of both $F$ and $P_{{\rm error}}$ Bob wishes to optimize, different values of $\eta$ may be optimum. Suppose, for example, that Bob would have to pay Alice a fine of \$1 if she figures out that his measurement outcome was wrong (she does know what outcome Bob ``should'' obtain), and he would have to pay the same fine if instead she projects onto a Bell state other than the state $|\Psi^+\rangle$. In that case, he would like to minimize the expectation value of his fine $f$ (also called the cost function), in units of \$1,
\begin{equation}\label{f}
\langle f\rangle=1-F+P_{{\rm error}}=1-(D_{{\rm rev}}+D_{{\rm rel}})/2.
\end{equation}
This implies he should choose $\eta$ such that $c(\eta)=\sqrt{2}/2$, so that the angle $\theta=\pi/4$, and hence $D_{{\rm rev}}=D_{{\rm rel}}=\sqrt{2}/2=0.707\ldots$ All this implies $\langle f\rangle_{\min}=1-\sqrt{2}/2=0.292\ldots$ In Figure~\ref{Fine} we plot  as the solid curve the expectation value of Bob's fine vs $D_{{\rm rel}}$ (one observer only).
\subsection{Multiple observers, eavesdroppers, and the environment}
An important requirement for a ``good'' or reliable measurement is that several independent observers should agree about the measurement outcome \cite{zurek}.
We can easily introduce multiple observers $k=2\ldots M$, in addition to Bob, each of whom taps off a fraction $\eta_k$ with $\sum_k\eta_k\leq1-\eta$ of the initial coherent-state amplitude of the counter. Each such observer would be modeled as an additional probe oscillator coupled to the counter as in Figure~\ref{domino}. Each observer $k$ gains information about the qubit state, given by expressions similar to (\ref{info}), determined only by the coupling strength $\eta_k$. 

On the other hand, the degree of reversibility decreases with each newly added observer $k$ who chooses a coupling strength $\eta_k> 0$. Thus, for each observer $k$ there is a different tradeoff relation between the individual degree of reliability of their own measurement and the overall degree of reversibility. 
If we denote by $\eta_T$ the sum of all the $\eta_k$ and $\eta$, that is, $\eta_T=\eta+\sum_k\eta_k$, then we have
\begin{equation}
D_{{\rm rev}}=(c_0)^{\eta_T},
\end{equation}
whereas Bob's degree of reliability is 
\begin{equation}
D_{{\rm rel}}=\sqrt{1-c(\eta)^2}.
\end{equation}
Hence we now have Bob's tradeoff relation, in the presence of other observers,
\begin{equation}\label{K}
KD_{{\rm rev}}^2+D_{{\rm rel}}^2=1.
\end{equation}
where we introduced a parameter $K$ according to
\begin{equation}\label{K2}
K=\frac{1}{c_0^{2\tilde{\eta}}},
\end{equation}
where $\tilde{\eta}$ is the sum of all $\eta_k$ of the  observers other than Bob, $\tilde{\eta}=\sum_k\eta_k$. 

In Figure~\ref{Fine} we plot the expectation value of Bob's fine in the presence of other observers, for several values of $K$.
It shows that Bob's fine will only increase with each addition of a new observer. Moreover, in order to minimize his fine, Bob will need to gain more and more information himself. This will be subject to the constraint that $\eta_T\leq 1$.

In Figure~\ref{rr} we plot the reversibility/reliability tradeoff, $D_{{\rm rev}}$ vs $D_{{\rm rel}}$ as determined by Eq.~(\ref{K}), for different values of  $K$. Whereas the degree of reliability can still reach the value 1, the maximum possible degree of reversibility decreases with increasing $K$  as $1/\sqrt{K}$.

As long as $N\epsilon$ is so large that $\sqrt{\eta_k}N\epsilon\gg 1$ even for small values of $\eta_k$, there may be many observers all independently obtaining reliable information about the qubit state \cite{zurek2}. That is, with very high probability (since $P_{{\rm error}}\ll 1$), each observer obtains the {\em same} measurement outcome. More precisely, suppose that $c_0\gg 1$, and assume we have $M$ observers in total, all having at their disposal a fraction $\eta_k=\eta=1/M$. Suppose also we wish the error probability of each observer to be limited to $P_{{\rm error}}\leq p$, for some small $p\ll 1$. Then it is easy to see that the number of observers must obey
\begin{equation}
M\leq \frac{2\log(1/c_0)}{\log(1/4p)}\equiv M_p.
\end{equation}
The number $M_p$ is directly related to the redundancy parameter $R_\delta$ of Refs.~\cite{zurek2}: it tells one
how redundantly information about the qubit state is stored in the many probes.

Finally, we note that the meaning of the coefficients $\eta_k$ for $k\geq 2$ may be different in different scenarios: for instance, in a cryptographic context, Alice and Bob may be the legitimate participants in a protocol, whereas an eavesdropper attempting to gain information can be characterized by a parameter $\eta_2$.
Or, in a more general context, unwanted interactions with the ``environment'' (i.e., any system not under control of Alice and Bob) may be characterized by parameters $\eta_k$ as well. In the latter case, the negative effects of those undesirable interactions are typically summarized as ``decoherence'' \cite{zurek}.
\subsection{Information and reliability}
So far, we have used the term ``information'' somewhat loosely to mean the same thing as reliability, in the sense that a reliable measurement provides Bob accurate information. Here we make the connection between the two concepts precise.

\begin{figure}
\includegraphics[width=4.5in]{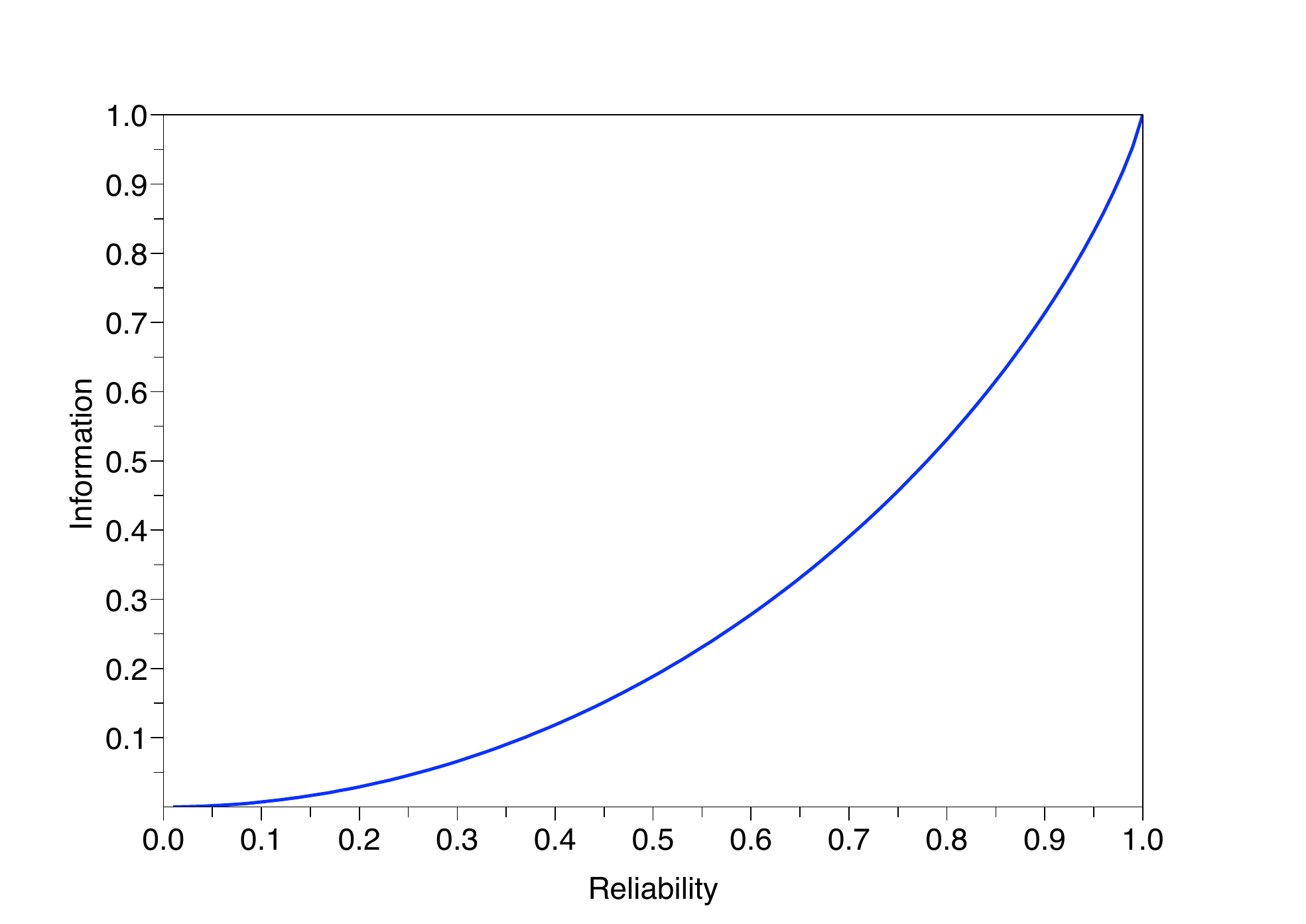}
  \caption{Mutual information $I$ versus degree of reliability
  $D_{{\rm rel}}$.  Both quantities quantify the correlation between Bob's measurement outcome and Alice's qubit $A$. See text for further details.}\label{ID}
 \end{figure}

In order to quantify the information that Bob could gain about the spin of qubit $A$  by trying to distinguish the coherent states $|-N\epsilon\rangle$ and $|+N\epsilon\rangle$ of Bob's probe,
we calculate the mutual information,
\begin{equation}
I(X;Y)=\sum_{x,y}P_{XY}(x,y)\log\frac{P_{XY}(x,y)}{P_X(x)P_Y(y)},
\end{equation}
where in our case $X$ refers to the qubit and $x$ takes the values $\uparrow$ and $\downarrow$, while $Y$ refers to the probe, and $y$ is assumed to take the values $+$ or $-$. Given the error probability
$P_{{\rm error}}$ of Eq.~(\ref{Perror}), we have
\begin{eqnarray}
P_{XY}(\uparrow,+)&=&P_{XY}(\downarrow,-)=(1-P_{{\rm error}})/2;\nonumber\\
P_{XY}(\downarrow,+)&=&P_{XY}(\uparrow,-)=P_{{\rm error}}/2,
\end{eqnarray}
and $P_X(\uparrow)=P_X(\downarrow)=P_Y(+)=P_Y(-)=1/2$.
All of this implies
\begin{equation}\label{info}
I(X;Y)=1-S(P_{{\rm error}}),
\end{equation}
with $S(P)=-P\log_2P-(1-P)\log_2(1-P)$ being the Shannon entropy function.
We plot $I$ as a function of $D_{{\rm rel}}$ in Figure~\ref{ID}.
 It shows that the mutual information is a monotonically increasing function of the degree of reliability, justifying our intuition that the two concepts are really the same in our case, up to a simple transformation.
In particular, we have
$I=1$ when $D_{{\rm rel}}=1$ (i.e., when $P_{{\rm error}}=0$) and 
$I=0$ when $D_{{\rm rel}}=0$ (i.e., when $P_{{\rm error}}=1/2$). 

\section{Conclusions}
The main result we want to emphasize is that there exists a constraining relation between the reliability of a quantum measurement and the extent to which the measurement process is (in principle) reversible, even in the absence of any external environment other than the fundamental information recording device. That is, if the person (Bob) making the measurement is cognizant of the result, meaning the result is stored in his brain, and if he subsequently refuses to allow his brain to be erased (unitarily reversed), then the measurement is irreversible (and non unitary). The very act of Bob deciding to remember permanently the result is sufficient to make the measurement irreversible, meaning that the measured quantum system cannot be restored to its original state by any unitary operation that is universal (not depending on  knowledge of the initial state).  Furthermore, the greater the information that is gained by Bob (that is, the more reliable the measurement is), the less reversible the measurement dynamics become. We don't claim these to be fundamentally new results, although we have not seen them verified in the literature in such simple terms.   

How do our results relate to the considerations by Poincar\'e of
the relation between probability, knowledge, and reversibility? He
pointed out that small errors in specifying final conditions of a
system after it underwent classical Hamiltonian evolution would
prevent the system from perfectly retracing its evolution under the
same Hamiltonian time-reversed. In our quantum case, letÕs say that
Bob wishes to perfectly reverse the evolution of the detector system
and qubit, along with his probe (that is, he chooses not to make a
projective measurement on his probe). He can do this reversal
perfectly, in principle, only if there are not other observers who
hoard some information in their probes for the same counter that Bob
is studying. That is, other observers necessarily change the state of
the system, making it less amenable to reversal from Bob's viewpoint. 

We could have imagined that the reversibility might be degraded exponentially
as a function of the amount of information gained in the measurement,
in loose analogy with the extreme sensitivity of the classical
reversibility of chaotic systems to small changes of final conditions.
This turns out not to be the case, and in hindsight this is not
surprising because quantum theory contains no chaos in the classical
sense of hypersensitivity to final (or initial) conditions on the state (although hypersensitivity
to small perturbations to the Hamiltonian does occur in quantum systems \cite{peresc,chaos,chaos2}). 
Rather, in our case both reversibility and information gain depend exponentially on the parameter $\eta$ that characterizes the strength of the measurement, such that the dependence of reversibility on information gain is polynomial, not exponential.

\section*{Acknowledgments}
The work of M.R. was supported by NSF grants PHY-AMOP-0456974 and PHY-PIF-0554842. 

\end{document}